\documentclass[11pt]{amsart}
\usepackage{vmargin} 
\usepackage{graphicx}
\usepackage{color} 
\setmargrb{1.25in}{1in}{1in}{1in}

\newlength{\graphicsheight}
\newcommand*{\schr}{Schr\"{o}dinger}
\newcommand*{\norm}[1]{\|#1\|} 
\let\grad\nabla
\newcommand*{\lap}{\Delta}
\newcommand*{\Ham}{\ensuremath{\mathcal{H}}}
\newcommand*{\N}{\mathcal{N}}
\newcommand*{\x}{\mathbf{x}}
\newcommand*{\setparskipindent}{\parskip 0.7ex \parindent 0pt} \setparskipindent
\newcommand*{\dstrat}{\circ\kern-0.06em d}
\newcommand*{\dstratdisplay}{\circ\kern-0.22em d}
\newcommand*{\dint}[1]{\int\kern-.7em\int #1 \,dx dy}
\newcommand*{\dintinline}[1]{\int\kern-.5em\int #1 \,dx dy}
\newcommand*{\p}{\mathbf{p}}

\begin{document}
\title
[Wave energy localization by self-focusing in large molecular structures]
{Wave energy localization by self-focusing in large molecular structures:
a damped stochastic discrete nonlinear Schr\"{o}dinger equation model}
\author{Brenton LeMesurier \and Barron Whitehead}
\address{Department of Mathematics, College of Charleston, Charleston SC}
\email{lemesurierb@cofc.edu}
\date{\today}

\begin{abstract}
Wave self-focusing in molecular systems subject to thermal effects, such as thin molecular films and long biomolecules, can be modeled by stochastic versions of the Discrete Self-Trapping equation of Eilbeck, Lomdahl and Scott, and this can be approximated by continuum limits in the form of stochastic nonlinear {\schr} equations.

Previous studies directed at the SNLS approximations have indicated that the self-focusing of wave energy to highly localized states can be inhibited by phase noise (modeling thermal effects)
and can be restored by phase damping (modeling heat radiation).

Here it is shown that the continuum limit is probably ill-posed in the presence of spatially uncorrelated noise, at least with little or no damping, so that discrete models need to be addressed directly.
Also, as has been noted by other authors, omission of damping produces highly unphysical results.

Numerical results are presented here for the first time for the discrete models including the highly nonlinear damping term, and new numerical methods are introduced for this purpose.

Previous conjectures are in general confirmed, and the damping is shown to strongly stabilize the highly localized states of the discrete models.
It appears that the previously noted inhibition of nonlinear wave phenomena by noise is an artifact of modeling that includes the effects of heat, but not of heat loss. 
\end{abstract}


\maketitle

\subsection*{Foreword}
This is a revised version of the material presented in several talks, given at the FPU+50 conference in Rouen, June 2005, at the University of Arizona, in September and November 2005, and at the University of New Mexico in October 2005.

\section{Introduction}

Combinations of mildly nonlinear wave motion in molecular structures with localized excitation modes can lead to localization of wave energy.
Perhaps the early mathematical example was Davydov's soliton theory modeling \emph{exciton waves} in protein molecules interacting with localized \emph{phonons}, vibrations at CO bonds, where a continuum limit gave the integrable one dimensional focusing cubic Nonlinear {\schr} (NLS) equation \cite{Davydov1971,DavydovK1973,Davydov1979}.
This was further developed by various authors including Scott \cite{Scott1982}, and extended to vibrations in other molecular systems such as crystalline acetanilide and in smaller molecules such as benzene \cite{EilbeckLS1984,EilbeckLS1985}.
Approximations that eliminate the fastest internal vibration modes again lead to systems that are discrete counterparts of the 1D focusing cubic NLS, or coupled systems of such.

Two dimensional molecular structures such as Scheibe aggregates \cite{BucherKuhn1970} lead to similar mathematical models related to the 2D focusing cubic NLS \cite{HuthGV1989,BartnickT1993}.

Stochastic terms are a natural refinement, modeling effects such as random spatial variations in the medium (\emph{fixed pattern noise}: time independent) or thermal agitation (\emph{white noise}: temporally uncorrelated) \cite{BangCIRG1994}.
Nonlinear optics has also produced continuum or semi-discrete examples including intense CW lasers and multi-cored optical fibers with random imperfections in the medium or in the strength of the coupling between signals in the different cores or different propagation modes.

However, noise without balancing losses can lead to an unphysical degree of spatial disorder or \emph{thermal runaway}, impeding wave propagation in contradiction to experimental observations \cite{ChristiansenGJRY1996}.
Even worse, the obvious continuum limits give Stochastic NLS (SNLS) equations which seem to be well-posed only when the noise has adequate spatial correlation \cite{deBouardDeB1999}, which is not necessarily consistent with the length scales in the molecular systems.
In other words, noise can destroy the spatial smoothness needed to justify a continuum limit.

More realistic modeling thus requires damping losses, to make possible attainment of thermal equilibrium, solutions with enough spatial smoothness to sustain traveling waves, and perhaps to justify a continuum limit model.
This leads \cite{ChristiansenGJRY1996} to the Damped Stochastic Discrete Nonlinear {\schr} equation (DSDNLS)
\begin{equation}\label{eq:DSDNLS}
i \frac{d\Psi_n}{dt} + \sum_{m} J_{nm}\Psi_m + |\Psi_n|^{2\sigma}\Psi_n
+ \gamma \Psi_n \frac{dW_n}{dt}
- \lambda \Psi_n \frac{d}{dt}(|\Psi_n(t)|^{2\sigma})
= 0,
\end{equation}
with $\sigma=1$ giving the cubic case typical in physical applications.
This is the main object of this study, but possible continuum limits will also be discussed briefly.
The most obvious is the Damped Stochastic Nonlinear {\schr} equation (DSNLS)
\begin{equation} \label{eq:DSNLS}
i \frac{\partial \psi}{\partial t} + \lap\psi + |\psi|^{2\sigma}\psi
{ + \gamma \psi \frac{\partial w}{\partial t}} - \lambda \psi \frac{\partial (|\psi|^{2\sigma})}{\partial t} = 0,
\end{equation}
derived and studied in \cite{ChristiansenGJRY1996}.
Here $W_n(t;\omega)$ and $w(\x,t;\omega)$ are noise processes with $\omega$ labeling realizations,  and the physical meanings of the variables and parameters will be explained in the next section.

The next section surveys the background for this study: physical origins, mathematical modeling, and previous results, some theoretical, but mostly numerical simulations.
Section~\ref{sect:numerical-methods} decribes the numerical methods used, and
Section~\ref{sect:numerical results} presents numerical results, primarily for the surrogate case of a one dimensional structure with quintic nonlinearity $\sigma=2$.
Then continuum limits are discussed in Section~\ref{sect:continuum}, with some suggestions as to how to overcome problems with previous approaches, followed by conclusions and a discussion of directions for further work.

\section{Background}
\label{sect:background}
We give here some details of a two dimensional example from chemistry and biochemistry: modeling \emph{Scheibe aggregates}, a class of highly ordered thin films of molecules coupled by dipole interactions predominantly within a plane: an essentially two dimensional wave medium.
However much of the modeling is also relevant to a variety of other molecular systems such as the essentially one dimensional protein models discussed above; see \cite{Davydov1979,Scott1982} and references therein.

\subsection{Scheibe aggregates: highly ordered thin molecular films}

Scheibe aggregates are highly regular arrangements of molecules in thin films, sometimes a single molecule thick, or with only weak interaction between neighboring layers of molecules.
These structures have important biological roles such as in photo-chemical reactions, and one laboratory example is the cyanine dye Scheibe aggregates first studied by B\"{u}cher, Kuhn, M\"{o}bius et al \cite{BucherKuhn1970,MobiusKuhn1988}.
They establish an arrangement of the molecules in a single layer ``brick-wall'' lattice, with the dominant dipole interactions being those with six nearest neighboring molecules, arranged in a hexagonal array of approximately dihedral $D_2$ symmetry (half turns and reflection in two perpendicular axes).

The molecules also have internal excitation states, which are coupled to the excitons and are also subject to thermal effects: random external forces due to collisions with molecules from outside the thin film.

\subsection{An exciton-phonon system with phase noise and damping}

Such thin films can be modeled as an exciton-phonon system with noise and damping acting on the internal modes as described by Bang, Christiansen, If, Rasmussen and Gaididei in \cite{BangCIRG1994}, which adds damping to the purely quantum mechanical modeling of Bartnick and Tuszy\'{n}ski \cite{BartnickT1993}:
\begin{gather}
\label{eq:excitons}
i \hbar \frac{d\Psi_n}{dt} + \sum_{m \neq n} J_{nm}\Psi_m + \chi u_n \Psi_n =  0
\\
\label{eq:phonons}
M \frac{d^2u_n}{dt^2} + M\lambda \frac{d u_n}{dt} + M\Omega_0^2 u_n - \chi |\Psi_n|^2
=  \gamma
\frac{dB_n}{dt}
\end{gather}
where
\\
$\Psi_n(t)$ is the exciton wave at location $n$,
\\
$u_n(t)$ is the elastic degree of freedom of the molecule at location $n$,
\\
$J_{nm}$ is the dipole-dipole interaction energy,
\\
$\chi$ is the exciton-phonon coupling constant,
\\
$dB_n/dt$ is random external forces, the formal time derivative of an independent Wiener process (Brownian motion) at each node,
\\
$\gamma$ is the strength of the random external forces,
\\
$\lambda$ is the damping coefficient,
\\
$M$ is the molecular mass, and
\\
$\Omega_0$ is the Einstein frequency of each oscillator.

These equations conserves the energy $\N = \sum_n |\Psi_n|^2$ under the Stratonovic interpretation of the stochastic term as discussed below, at least on a fully infinite lattice or with suitable boundary conditions such as periodic.

The form of Eq's~\eqref{eq:excitons},\eqref{eq:phonons} also covers a wide range of other applications such as the one dimensional protein molecule models mentioned above: the indices will often be taken below to simply enumerate a collection of \emph{nodes}, with details such as spatial relationships between locations encoded in the coupling terms $J_{nm}$.
For further mathematical flexibility, the coupling term will henceforth be written as a general power law
$\chi |\Psi_n|^{2\sigma}$, though all physical models we know of have the cubic nonlinearity $\sigma=1$.
The underlying spatial dimension will be denoted by $D$, so $D=2$ in the model above.

Without noise and damping, the system in \eqref{eq:excitons},\eqref{eq:phonons} is Hamiltonian, giving a second conserved quantity
\[
\Ham = -\sum_{n,m \neq n} J_{nm}\Psi_n \overline\Psi_m - \frac{2\chi}{1+\sigma} \sum_n u_n |\Psi_n|^{2\sigma} + \frac{M}{2} \sum_n ( \dot{u_n}^2 + \Omega_0^2u_n^2 ).
\]

\subsection{Eliminating the phonon terms \boldmath $u_n$}

The  phonon terms $u_n$ can be eliminated using the variation of parameters formula which gives an integral expression for $u_n$ in terms of $\Psi_n$ \cite{ChristiansenGJRY1996}.
To eliminate the resulting time integral and initial data transients, one must restrict to times $\lambda t>>1$ and make the slowly varying envelope approximation: that the exciton intensity $|\Psi_n(t)|^2$ is slowly varying relative to the phonon frequency $\Omega_0$.
Thus, the presence of damping ($\lambda > 0$) is essential.

The reduced system is
\begin{eqnarray}\label{eq:DSDNLS1}
0 &=& i \hbar\frac{d\Psi_n}{dt} + \sum_{m \neq n} J_{nm}\Psi_m + V|\Psi_n|^{2\sigma}\Psi_n
\\
\nonumber
&&
+ \frac{\gamma \chi}{M \hbar \Omega} \Psi_n \frac{dW_n}{dt}
- \frac{\lambda V}{\Omega_0^2}\frac{d}{dt}[|\Psi_n(t)|^2] \Psi_n,
\mbox{ where}
\\
\nonumber
V &=& \chi^2/M\Omega_0^2, \quad
\Omega^2 = \Omega_0^2 - (\lambda/2)^2, \mbox{ and}
\\
\label{eq:sigma-from-eta}
\frac{dW_n}{dt} &=&
\lambda \int_0^t e^{-\lambda s/2} \sin(\Omega s) \dot B_n(t-s) ds.
\end{eqnarray}
Note that the new noise processes $W_n$ are temporally correlated, except in certain limits such as strong damping.

\subsection{Rephrasing as a Damped Stochastic Discrete Nonlinear {\schr} Equation}

If the coupling $J_{nm}$ is homogeneous, in that the quantities
\[
J_{nn} := -\sum_{m \ne n} J_{nm}
\]
have a common value $J_0$, then the $\Psi_n$ have a common average phase evolution $e^{tJ_0/\hbar}$.
This can be removed by adding $J_{nn}$ to each coupling sum,
and with some rescalings including
${\gamma \chi}/{(M \hbar^2 \Omega)}  \rightarrow \gamma$,
${\lambda}/{(\hbar \Omega_0^2)} \rightarrow \lambda$,
one gets the Damped Stochastic Discrete Nonlinear {\schr} equation Eq.~\eqref{eq:DSDNLS}.

Even if the $J_{nn}$ are not all equal, one can use their average value $\bar{J}$ as the phase shift, and absorb the differences $J_{nn}-\bar{J}$ into the noise coefficients as fixed pattern noise.

The energy $\N = \sum_n |\Psi_n|^2$ is still conserved, and without noise or damping the system still has a conserved Hamiltonian,
\[
\Ham = - \sum_{n,m} J_{nm} \Psi_n \overline{\Psi}_m  - \frac{1}{1+\sigma} \sum_n |\Psi_n|^{2(1+\sigma)}.
\]

Simple examples are uniform nearest neighbor interaction on a line [1D] or square lattice [2D] of spacing $l$ with all the non-zero $J_{nm}$ having the same value, $J/l^2$.
The coupling term is then a $J$ times the standard three point second derivative [1D] or five point discrete laplacian [2D], and the equation is a discretization of the Damped Stochastic Nonlinear {\schr} equation \eqref{eq:DSNLS}.
Also, the Hamiltonian for the case of no noise or damping is the natural discretization using simple forward difference quotients for the gradient terms of the Hamiltonian for NLS,
\[
\Ham = \int \norm{\grad \psi}^2  - \frac{1}{1+\sigma} |\psi|^{2(1+\sigma)} \;d\x.
\]

It can be useful in places to think of the ODE systems in relation to the familiar NLS equation, but a more careful consideration of continuum limit approximations is needed, as discussed in section \ref{sect:continuum}.

\subsection{Removing the extra time derivative term, and Stratonovic differential form}

For some purposes, the time derivative should be eliminated from the damping term.
Also, the rigorous mathematical formulation must be in terms of stochastic integrals and differentials, and in order to conserve the energy $\N$, products involving stochastic terms must be interpreted in the Stratonovic sense.
(Loosely, the Stratonovic integral is defined as the limit of midpoint rule (or trapezoid rule) approximations, whereas the It\^{o} integral is the limit of left-hand end point Riemann sums: see \cite{Oksendahl2000} for details.)
Solving for $d\Psi_n/dt$, substituting into the other time derivative leads to the stochastic differential form
\begin{equation} \label{eq:DSDNLS-diff}
i d\Psi_n + \left[ \sum_{m} J_{nm}\Psi_m + |\Psi_n|^{2\sigma}\Psi_n
+ 2 \lambda \Psi_n\mbox{Im} \left( \overline{\Psi}_n\sum_{m} J_{nm}\Psi_m \right) \right] dt
+ \gamma \Psi_n {\dstrat W_n}
= 0
\end{equation}
with $\dstrat$ denoting the Stratonovic differential.

\subsubsection*{Why not convert to It\^{o} integral form?}

Any system of stochastic ODE's in terms of the Stratonovic integral can be replaced by an equivalent system which gives the same solution under the It\^{o} interpretation, by replacing the above Stratonovic differential by the corresponding It\^{o} differential plus a correction term \cite{Oksendahl2000}, and the It\^{o} form is far more amenable to analysis such as existence and uniqueness proofs
In the current case, this gives
\begin{equation}\label{eq:Ito-correction}
d\Psi_n = \mbox{[as before]} + i \gamma \Psi_n {d W_n} - \frac{\lambda^2}{2} \Psi_n
\end{equation}

However, this form is undesirable for current purposes, particularly numerical simulations.
The new term adds rapid exponential decay, destroying the manifestly conservative form.
This reflects the fact that the It\^{o} differential term itself generates rapid exponential growth of individual realizations, related to the fact that the ensemble average of It\^{o} solutions satisfy the underlying noise-free equation, and so conserves the energy $\N$.

This prevents the use of time discretizations which inherently conserve energy $\N$; such conservative discretizations are used here for the Stratonovic form.

Also, the  It\^{o} form has no continuum limit with spatially uncorrelated noise.
This might reflect the conjectured lack of existence of solutions to such continuum limits, even when the limit formally exists for the Stratonovic form.

\subsection{Wave self-focusing and energy localization in SNLS}
\label{subsect:background, localization in snls}

In the continuum model of 2D NLS, self-focusing can lead to the formation of single point singularities, sometimes called \emph{wave collapse}.
The proof of this for the NLS is based on a variance argument, which has been extended to various cases of Stochastic NLS by Gaididei and Christiansen \cite{GaidideiChristiansen1998}, Debussche and Di Menza \cite{DeBusscheDiMenza2002}, and Fannjiang \cite{Fannjiang2005}, though all require noise that is sufficiently correlated in space and uncorrelated in time.
The last mentioned author's results are as follows.

For data of sufficiently rapid decay at infinity, the pulse width can be measured by its spatial variance
\[
V(t) = \int |\psi|^2 \norm{\x}^2 d\x.\]
In the critical case of  2D cubic (and more generally, $\sigma D=2$), the ensemble average $\langle V \rangle$ is related to the ensemble average $\langle \Ham \rangle$ of the Hamiltonian
by
\begin{equation}\label{eq:variance-deriv-snls}
\frac{d^2 \langle V \rangle}{d t^2} = 8 \langle \Ham \rangle,
\end{equation}
and in turn noise modifies conservation of  the Hamiltonian to
\begin{equation}\label{eq:ham-deriv-snls}
\frac{d \langle \Ham \rangle}{dt} = R := \frac{1}{2}\int \Phi(\p) \norm{\p}^2 d\p,
\end{equation}
where $\Phi(\p)$ is the power spectral density of the noise distribution.

For spatially uncorrelated noise, $\Phi(\p)$ is  a positive constant, so $R=\infty$: the formulas break down, but strongly suggest that SNLS has no solutions, at least in Sobolev space $H^1$.
For comparison to the discrete equations, for which no analogues of these formulas are known, note that a discretization of SNLS with grid spacing $l$ effectively has noise of correlation length scale $l$, and such correlation in SNLS gives
\begin{equation}\label{eq:R}
R=O\left(\frac{1}{l^{2+D}}\right).
\end{equation}

Without noise, $R=0$, $\langle V \rangle=V$ evolves quadratically, and $\Ham<0$ is sufficient condition for finite time singularity formation, as otherwise, $V$ would become negative.
Noise changes the evolution to the cubic
\begin{equation}\label{eq:variance-ev-snls}
\langle V \rangle = \langle V \rangle(0) + bt + 4 \langle \Ham \rangle(0) t^2 + 2R/3 t^3.
\end{equation}
Clearly, with weak enough noise, $\Ham(0)<0$ still leads to a prediction of negative $\langle V \rangle$ by some positive time $t_0$, so with positive probability, solutions must cease to exist before that time.
However, sufficiently large noise eliminates this necessity, and hints at global existence, and at dispersion with a cubic rate of variance growth.

Such formulas have not yet been extended to account for damping, but as seen below, there are hints of an additional negative term in $d\Ham/dt$.

\subsection{Wave self-focusing and energy localization in the discrete systems}
\label{subsect:background-localization-discrete}

With no noise or damping, numerical simulations of discrete counterparts of the NLS equation show phenomena analogous to wave collapse, even in the subcritical case of 1D cubic, where the PDE has some degree of self-focusing, but cannot develop singularities.
That is, solutions can have the energy concentrate until it is mostly at a single node (molecule), and then stay localized in a solution that seems to oscillate around a stable steady state.
This was first described by Davydov~\cite{Davydov1979} in models of protein molecules, and further analyzed and simulated by Eilbeck, Lomdahl and Scott~\cite{Scott1982,EilbeckLS1984,EilbeckLS1985}, who describe the phenomenon as \emph{self-trapping}.

For the discrete systems with noise, no formulas are known for the evolution of ensemble averages considered above, but from Eq.~\eqref{eq:DSDNLS} and alternative form \eqref{eq:DSDNLS-diff}
the Hamiltonian evolution for individual realizations satisfies
\begin{eqnarray}\label{eq:dHdt-dsdnls}
\frac{d\Ham}{dt}
&=& \sum_n \left\{\frac{d W_n}{dt} \frac{d}{dt}(|\Psi_n|^2) - \lambda \left[\frac{d}{dt}(|\Psi_n|^2)\right]^2 \right\}
\\
\nonumber
&=& \sum_n \left\{ \frac{d W_n}{dt} \mbox{Im} \left( \overline{\Psi}_n\sum_{m} J_{nm}\Psi_m \right)
- 4 \lambda \mbox{Im} \left( \overline{\Psi}_n\sum_{m} J_{nm}\Psi_m \right)^2 \right\}.
\end{eqnarray}

These no longer make it clear that noise causes the previously noted linear increase in $\Ham$, or corresponding growth in beam spatial variance or inhibition of wave collapse, but all these are still seen in numerical studies of discrete systems with spatially uncorrelated noise, including those below.
This is to be expected, since those discrete systems are effectively discretizations of SNLS with noise of spatial correlation on length scale comparable to the mesh spacing of the discretization.

The full 2D Stochastic NLS equation with spatially uncorrelated noise has been simulated by Bang, Christiansen, If, Rasmussen and Gaididei in\cite{BangCIRG1994}, and the 1D quintic case of the Stochastic NLS equation by DeBussche and Di Menza in \cite{DeBusscheDiMenza2002}.
In each case it is observed that  spatially uncorrelated noise above a certain threshold level prevents wave collapse.

The former paper indicates that this noise effect is too strong to match physical experiments:
noise levels so low as to correspond to temperatures of a few Kelvin are needed to reproduce behavior seen in experiments at far higher temperatures.
The likely cause is ``thermal runaway'' due to the absence of a mechanism for ``heat loss'', such as the damping term.

The latter authors also interpret simulations as showing that with spatially correlated noise (as is effectively imposed by a fixed spatial discretization), collapse can only be delayed, but will always occur.
They also offer a non-rigorous argument for this second conclusion.

However we come to a different conclusion below.
In \cite{DeBusscheDiMenza2002}, the authors judge that collapse has occurred whenever the maximum amplitude has grown by a factor of three at any one time, but this seems unreliable when noise is present, since then amplitude spikes this high at a single node can occur transiently, rather than as part of ongoing focusing events.
A more reliable criterion for numerical detection of wave collapse and energy localization is that the energy at a single node exceeds some substantial fraction of all energy, persistently for a significant interval of time.
In the simulations below, localization is manifested with a majority of all energy staying at one node until the end of the computer time, and thus persisting for at least as long as the rise time of the localization.

As to damping effects, Eq.~\eqref{eq:dHdt-dsdnls} indicates that damping has the opposite effect of noise, causing reduction of $\Ham$.
Combined with the expectation that \eqref{eq:variance-deriv-snls} still holds approximately for discretizations of SNLS, this suggests the possible return of wave collapse, in the discrete form of concentration of energy near a single node.

Christiansen et al \cite{ChristiansenGJRY1996} have done the only simulations known to the current authors of the model with damping of Eq.~\eqref{eq:DSNLS}.
They do this with further approximate by a small system of ODE's, first imposing radially symmetry and then using the method of \emph{collective coordinates}.
Such modeling has lead to some analytical results on self-focusing in the NLS equation, but with the stochastic terms, it still requires study primarily by simulation.

They start with simulations without damping, corroborating the above described observations about inhibition of wave collapse by sufficiently strong noise, leading instead to dispersion of initially concentrated wave energy.
With damping added, they observe that the effect of noise can be reversed, leading to collapse where with noise alone it would not occur, as suggested by Eq.~\eqref{eq:dHdt-dsdnls}.
Again there appears to be a threshold damping level for this to occur with given initial data and noise level.
\label{todo1}

With this survey done, we are ready to consider new numerical methods and results of simulations based on Eq.~\eqref{eq:DSDNLS}.

\section{Numerical methods: a variant of Chang and Xu's iterative trapezoid}
\label{sect:numerical-methods}

With noise but no damping, and with homogeneous coupling and periodic boundary conditions, a Fourier split-step method could be used, as was done by Bang et al \cite{BangCIRG1994} for SNLS.

Instead, an implicit time discretization based on fixed point iterative solution of the trapezoid rule is used, similar to one described and analyzed by Chang and Xu~\cite{ChangXu1986}.
It has several virtues:
\begin{itemize}
\item it satisfies the needed Stratonovic interpretation (no need for the It\^{o} correction term),
\item it conserves the exciton energy $\N$, and
\item with no noise or damping, it conserves the hamiltonian $\Ham$.
\end{itemize}
The main disadvantage is the need for iterative solution (so that conservation is no longer exact, but still highly accurate), and the fact that it is difficult to go beyond simple fixed point iteration due to the existence of coupled nonlinear terms, leading to the time step size restrictions typical of an explicit method rather than the underlying implicit method and a large number of iterations needed when noise effects are strong.
Since the coupling is strong relative to the other terms (as in the discretized NLS equations that these essentially are) the system is rather stiff, leading to the need for rather small time steps.

However, arguably the short spatial scales of the noise mean that the time step size limitations of simple iterative solution method are also the natural time scales of the smallest relevant features, so there might be little room for time step increases without  inaccurate handling of noise effects.

That is, the stiffness of the ODE systems is probably an essential time scale that must be preserved in the modeling, including the fully discrete model used for numerical solution.

\subsection{Trapezoid method: conservative time discretization}

Writing $\Psi_n^j$ for the approximations of $\Psi_n(t^j)$,
$\delta t = t^{j+1}-t^j$, and
$\delta W_n^j/\delta t$ for the approximation of $\sigma_n(t)$, constant on $t^j \leq t \leq t^{j+1}$,
the scheme used for the cubic case is
\begin{eqnarray} \label{eq:trapezoid}
0 &=& i \frac{\Psi_n^{j+1} - \Psi_n^j}{\delta t} + \sum_m J_{mn} \frac{\Psi_m^j + \Psi_m^{j+1}}{2}
\\
\nonumber
&+& \Bigg[ \gamma \frac{\delta W^j}{\delta t} + \frac{|\Psi_n^j|^2+|\Psi_n^{j+1}|^2}{2}
\\
\nonumber
&& + \lambda \mbox{Im}
\left( \overline{\Psi_n^j} \sum_m J_{mn} \Psi_m^j+\overline{\Psi_n^{j+1}}\sum_m J_{mn}\Psi_m^{j+1}\right) \Bigg]
\times \frac{\Psi_n^j + \Psi_n^{j+1}}{2}.
\end{eqnarray}
For the case studied so far of totally uncorrelated noise, the noise components $W_n^j$ are independent with normal distribution
$\delta W_n^j \sim N\left(0,{\delta t}/{l}\right),$
so that
${\delta W_n^j}/{\delta t} \sim N\left(0,{1}/(l\delta t)\right)$.
The scaling with $l$ and nearest coupling strength $J_{nm}=1/l^2$ are used for consistency with interpretation as the discretization of the DSNLS, Eq.~\eqref{eq:DSNLS}.

More generally, exact conservation of the Hamiltonian is achieved by discretizing the nonlinear term
$|\Psi_j|^{p-1}\Psi_j$
as
\[
\left(\frac{2}{p+1}\right)
\frac{|\Psi_n^{j+1}|^{p+1}-|\Psi_n^j|^{p+1}}{|\Psi_n^{j+1}|^2-|\Psi_n^j|^2}
\times \frac{\Psi_n^{j+1} + \Psi_n^j}{2}.
\]
Thus for quintic nonlinearity, the form for this term is not the familiar trapezoid approximation, but
\[
\frac{|\Psi_n^j|^4+|\Psi_n^j|^2|\Psi_n^{j+1}|^2+|\Psi_n^{j+1}|^4}{3}\frac{\Psi_n^j + \Psi_n^{j+1}}{2}.
\]

\subsection{Trapezoid method: iterative scheme}

The nonlinear implicit scheme above is solved using a simple fixed point iteration, eliminating implicit form even for the linear dipole coupling (discrete laplacian) terms, and thus avoiding simultaneous linear equation solving, contrary to the original Chang-Xu algorithm.
The reason is the nonlinearity and stiffness of the damping term and the formally unbounded noise terms, which lead in practice to similar ``explicit scheme'' time step size restriction $\delta t = O( l^2)$
even with implicit handling of the linear coupling term.

Writing $\Psi_n^{j,k}$ for the $k$-th iterate, the initial approximation used for the new time step is $\Psi_n^{j+1,0}=\Psi_{n,j}$, and
and subsequent iterates are given by solving the uncoupled linear equations
\begin{eqnarray}
\label{eq:trapezoid-iterative}
0 &=& i \frac{\Psi_n^{j+1,k+1} - \Psi_n^j}{\delta t} + \sum_m  J_{mn} \frac{\Psi_m^j + \Psi_m^{j+1,k}}{2}
\\
\nonumber
&+& \Bigg[\frac{|\Psi_n^j|^2+|\Psi_n^{j+1,k}|^2}{2}
+ \gamma\frac{\delta W_n^j}{\delta t}
\\
\nonumber
&+& \lambda \mbox{Im} \left( \overline{\Psi_n^j} \sum_m J_{mn} \Psi_m^j + \overline{\Psi_n^{j+1,k}} \sum_m J_{mn} \Psi_n^{j+1,k} \right) \Bigg]
\times \frac{\Psi_n^j + \Psi_n^{j+1,k+1}}{2}.
\end{eqnarray}

\subsection{Time step size}

The fixed point scheme has a worst case convergence rate of $K=\kappa\delta t$ with $\kappa=\max_n\sum_n|J_{nm}|/2$, so $\kappa=2/l^2$ for the three point discrete second derivative used in the 1D quintic case, giving a convergence condition $\delta t < 1/\kappa, = l^2/2$.
To minimize computational cost in the sense of minimizing expected iterations per unit time, the optimal choice of $K$ balances $O(1/K)$ time steps per unit time and the $O(\ln \epsilon/\ln K)$ iterations per time step needed to meet a given error tolerance $\epsilon$.

Minimization of $|\ln \epsilon/(K \ln K)|)$ gives $K=1/e$, which is in fact observed to be optimal with spatially uncorrelated noise and no damping, a case where it will be seen below that ``thermal runaway'' puts significant amount of signal in the shortest length scales, realizing the worst case convergence rate.
On the other hand, without noise or with damping, solutions are smoother and larger values $K \approx 0.5$ are most efficient, with $K<1$ always necessary for convergence.

\section{Numerical results}
\label{sect:numerical results}

For computational efficiency, simulations have been done with a single computational space dimension, using two different approaches to this reduction of dimension.

The main studies are done for the one dimensional quintic case $D=1$, $\sigma=2$, with this somewhat unnatural nonlinearity power used for the sake of remaining in the critical case for collapse in NLS.
Homogeneous nearest neighbor coupling is used, corresponding to discretizing NLS with the the standard three point discretization of the second derivative.
Nearest neighbor coupling makes sense in the physical models, since dipole interactions are very short range.
It also makes little sense to use higher order spatial discretizations of Stochastic NLS equations, because the noise eliminates the higher order smoothness needed to make such discretizations more accurate.
This case, without damping, was also studied by DeBussche and Di Menza \cite{DeBusscheDiMenza2002}, as discussed in section~\ref{subsect:background-localization-discrete} above.

The second reduction used is imposing radial symmetry on the two dimension cubic NLS ($D=2$, $\sigma=1$), and then again using standard three point discretization of spatial derivatives.
This allows comparison to the results of Bang, Christiansen et al \cite{BangCIRG1994,ChristiansenGJRY1996} also discussed above.

\subsection{1D lattice with quintic nonlinearity and homogeneous nearest neighbor coupling}
\label{subsect:results, 1D Q}

The initial data used in this section is always discretization of $\Psi_n(0)=\psi_0(x_n) = 1.1(1+\cos x_n)/2$ on a uniform periodic grid of $n_{nodes}$ equally spaced nodes in the period cell $[-\pi,\pi]$.
This is chosen to be close to the ``Townes soliton'' $R_0$ central to theory of NLS self-focusing, giving Hamiltonian $\Ham$ just slightly negative so as to ensure self-focusing in NLS, while having energy $\N= \N{\psi_0}$ just slightly above the minimum value $\N(R_0)$ needed for self-focusing to be possible (c.f. \cite{SulSul1999}).

The default parameters are $n_{nodes}=100$,
and noise strength $\gamma=0.04$, damping strength $\lambda=0.002$ when noise or damping are present at all,
with the choice of the latter two values explained below.
Also, results for a single ``standard noise realization'' are presented in many graphs, with corroboration by data from multiple realizations where appropriate.

In the figures, a systematic curve color coding is used.
Black is used only for initial data with functions of node index, and for averages over multiple realizations with functions of time.
The color sequence \textcolor{blue}{blue}, \textcolor{green}{green}, \textcolor{red}{red}, \textcolor{cyan}{cyan}, \textcolor{magenta}{magenta}, \textcolor{yellow}{yellow} is used both for later times with functions of node index, and for successive realizations with functions of time.

\subsubsection{Self-focusing and energy localization without noise or damping}
\label{subsubsect:results, 1D Q, no n,d}

\begin{figure}[htbp]
\includegraphics
[height=\graphicsheight]
{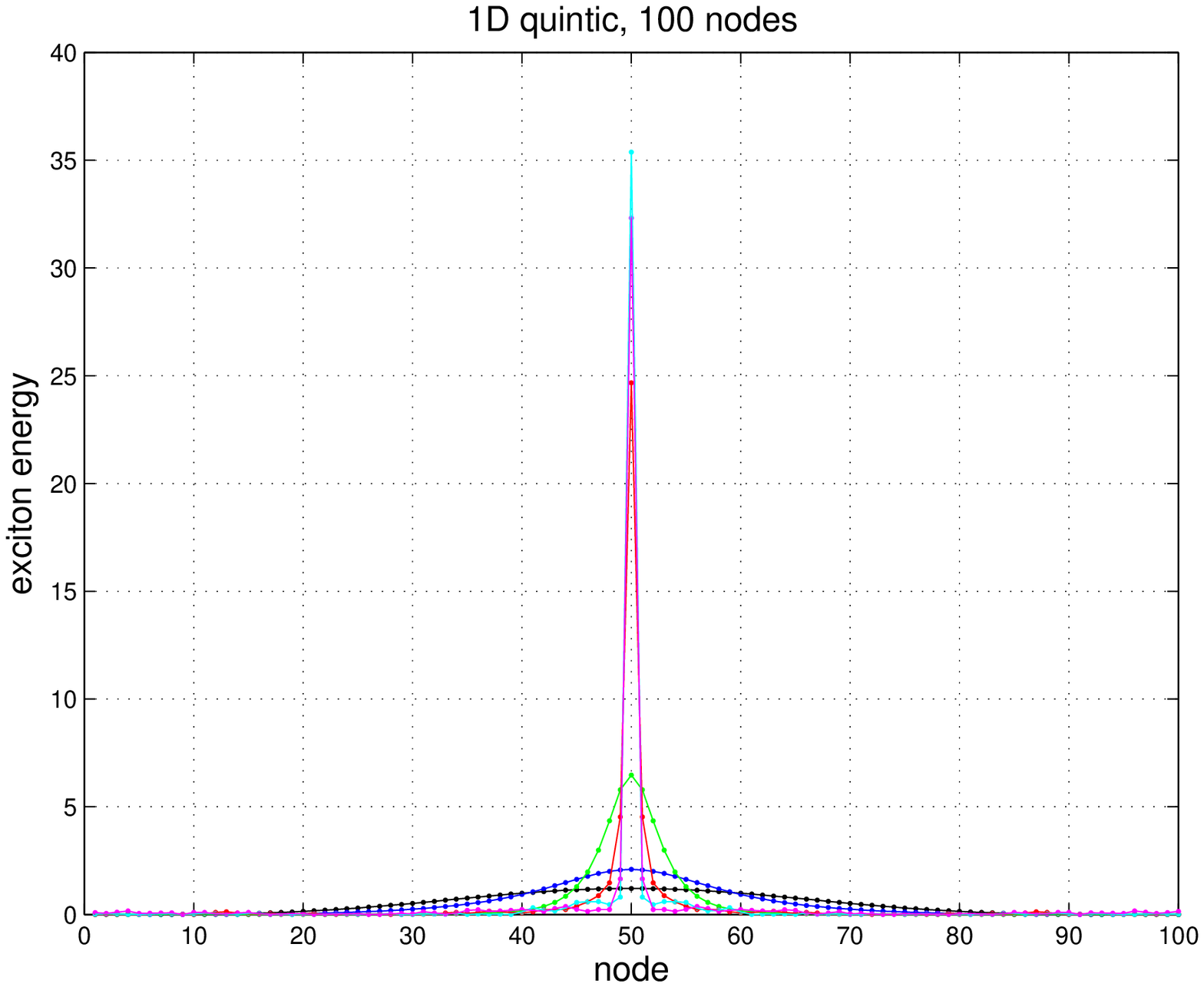}
\caption{Energy distribution for 1D quintic, no noise or damping.
Times and curve colors
$t=0, \textcolor{blue}{2}, \textcolor{green}{2.5}, \textcolor{red}{3}, \textcolor{cyan}{3.5}, \textcolor{magenta}{4}$
as noted in the text.}
\label{fig:dsdst1dq_etx_all}
\end{figure}

\begin{figure}[htbp]
\includegraphics
[height=\graphicsheight]
{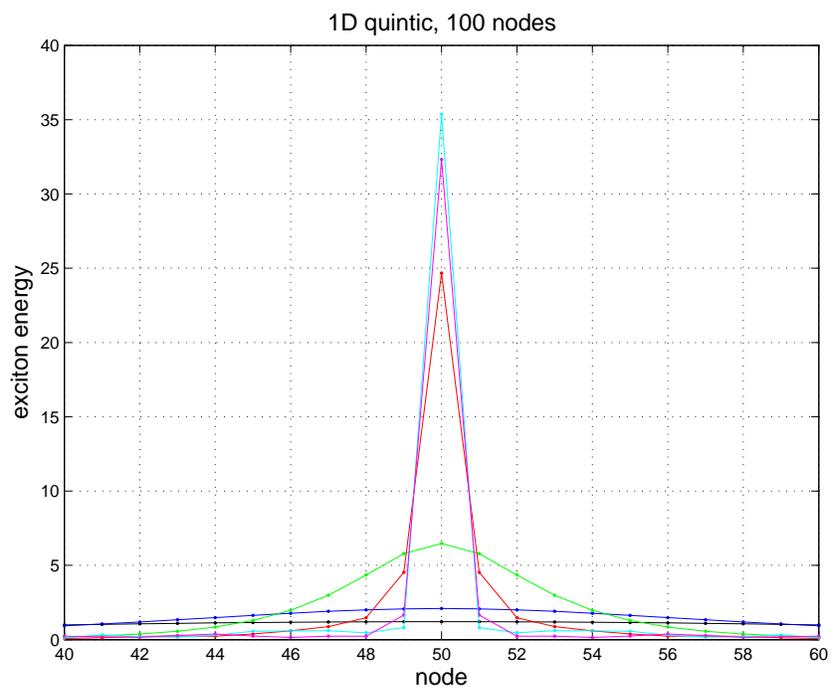}
\caption{
As in Fig.~\ref{fig:dsdst1dq_etx_all} but restricted to nodes near the self-trapping locus, showing energy persistently concentrated almost entirely at a single node.}
\label{fig:dsdst1dq_etx}
\end{figure}

The time evolution of Eq.~\eqref{eq:DSDNLS} in the 1D quintic case without any noise or damping is illustrated in Fig's~\ref{fig:dsdst1dq_etx_all} and \ref{fig:dsdst1dq_etx}, which show the distribution of energy $|\Psi_n(t)|^2$ amongst nodes, for various times $t$.
The first figure shows all 100 nodes, at three times before self-trapping occurs at
$t=t^* \approx 2.6$ and two times afterwards.
The second and all subsequent graphs of energy distribution are restricted to nodes near the self-trapping locus.

Note that once the energy at any node passes about 10, and certainly once it passes 20, energy is largely concentrated on only a few nodes so that the solution is no longer accurate as a discrete approximation of NLS.
Likewise, in all subsequent solutions with the current spacing of $l=50/\pi$, data past the time when the energy at any node first exceeds about 10 should only be considered as accurate for the spatially discrete models.
Indeed the maximum possible energy at one node is $\N/l$, which for the current initial data is approximately $45.3$.
Thus the values seen here of about 35 and above at a single node represent a clear majority of all energy concentrated at that node.

\begin{figure}[htbp]
\includegraphics
[height=\graphicsheight]
{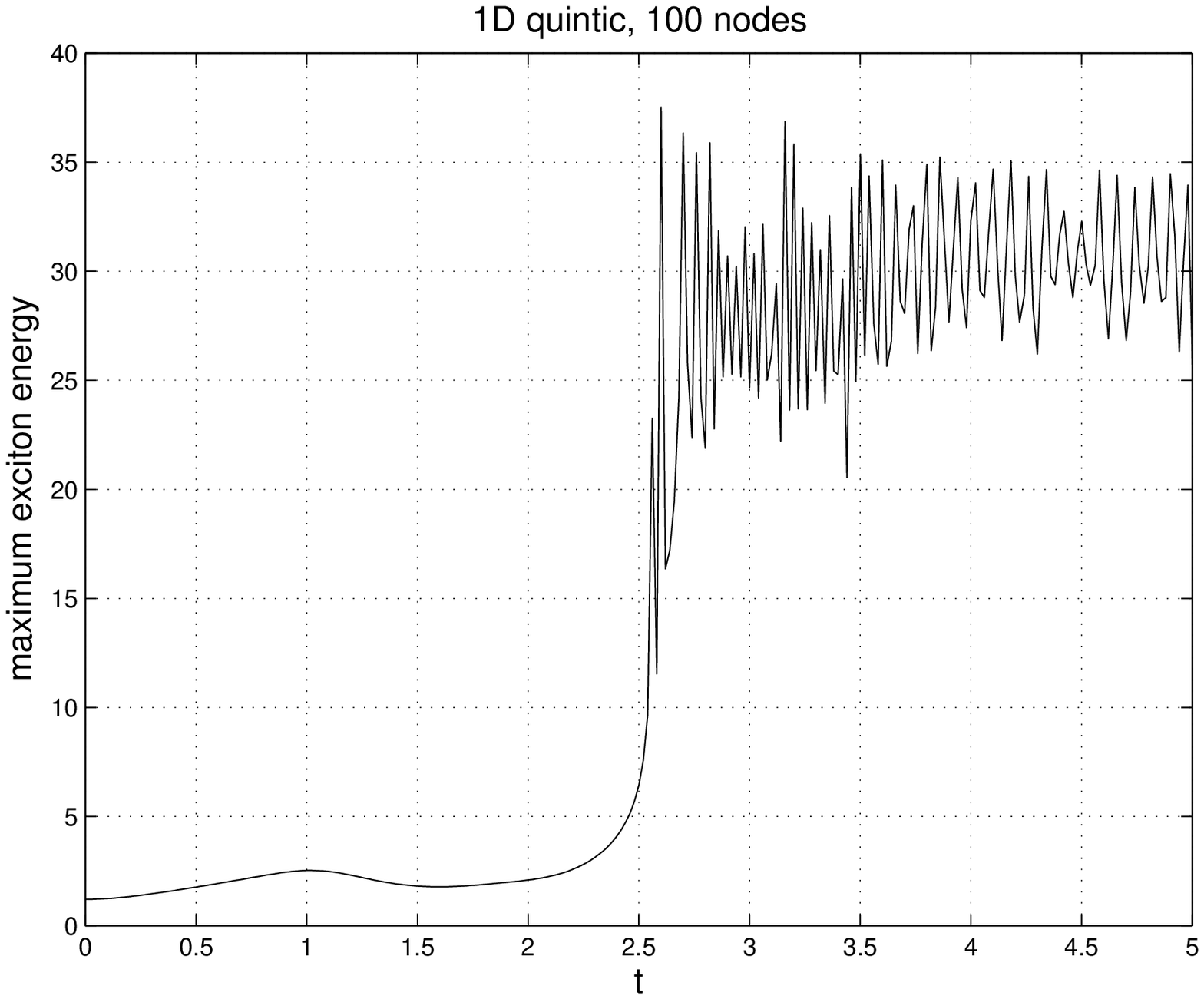}
\caption{Evolution of maximum single node energy for 1D quintic, no noise or damping.}
\label{fig:dnls_edmaxt}
\end{figure}

The evolution of the degree of energy localization is shown in Fig.~\ref{fig:dnls_edmaxt}, as measured by the maximum energy at any one node as a function of time.
Once energy localization has occurred, it is seen to persist, but with significant oscillations.
The oscillations fit with the idea of the solution entering a neighborhood of an orbitally stable stationary state that is a center: the existence of such localized stable center stationary states has been proven in the minimal case of $n_{nodes}=2$ by Eilbeck, Lomdahl and Scott~\cite{EilbeckLS1985}.

Note that this oscillatory behavior is purely a property of the discrete system, as it only sets in after the maximum single node energy becomes too high for the numerical solutions to be relevant to the related PDE models.

\subsubsection{Inhibition of self-focusing by sufficient noise, without damping}
\label{subsubsect:results, 1D Q, n, no d}

\begin{figure}[htbp]
\includegraphics
[height=\graphicsheight]
{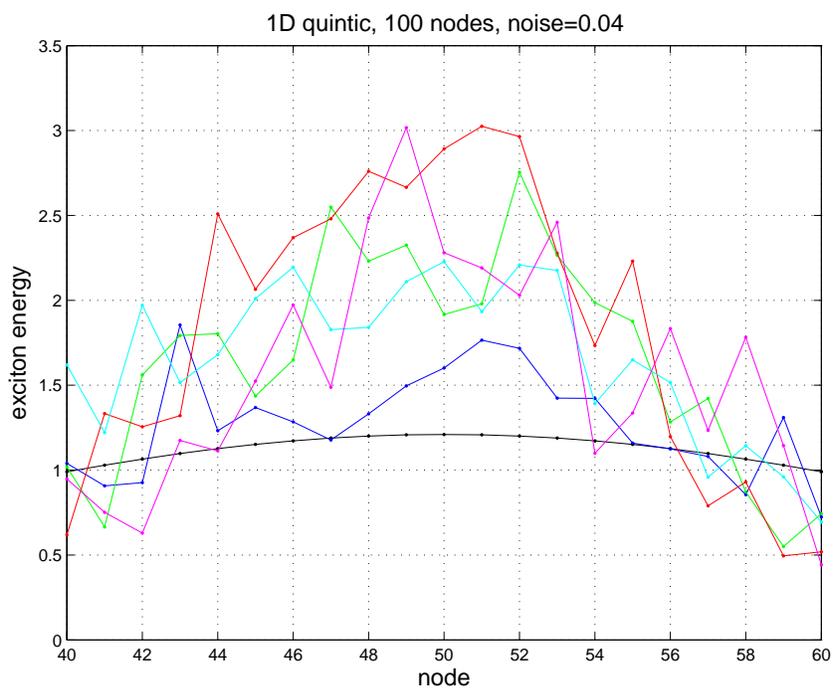}
\caption{Energy distribution for 1D quintic, noise $\gamma=0.04$, no damping.
Color-time labeling as in Fig.~\ref{fig:dsdst1dq_etx_all}.}
\label{fig:1dqnn_2}
\end{figure}

The results here are much as seen in the simulations by various previous authors discussed in
section~\ref{subsect:background-localization-discrete}.
With low levels of noise, up to $\gamma \approx 0.03$, focusing of energy to a single node followed by persistent localization still occurs.
Higher noise levels of $\gamma \geq 0.04$ inhibit self-focusing and energy localization, as shown in Fig.~\ref{fig:1dqnn_2} for the single standard noise realization, and confirmed by multiple realizations.

Another notable feature is the loss of the spatial smoothness that would be needed for continuum limit PDE modeling.
It seems likely that this spatial disorder has the effect of inhibiting exciton wave propagation, and that this is the mechanism which prevents energy localization, by preventing the needed energy flux.

\begin{figure}[htbp]
\includegraphics
[height=\graphicsheight]
{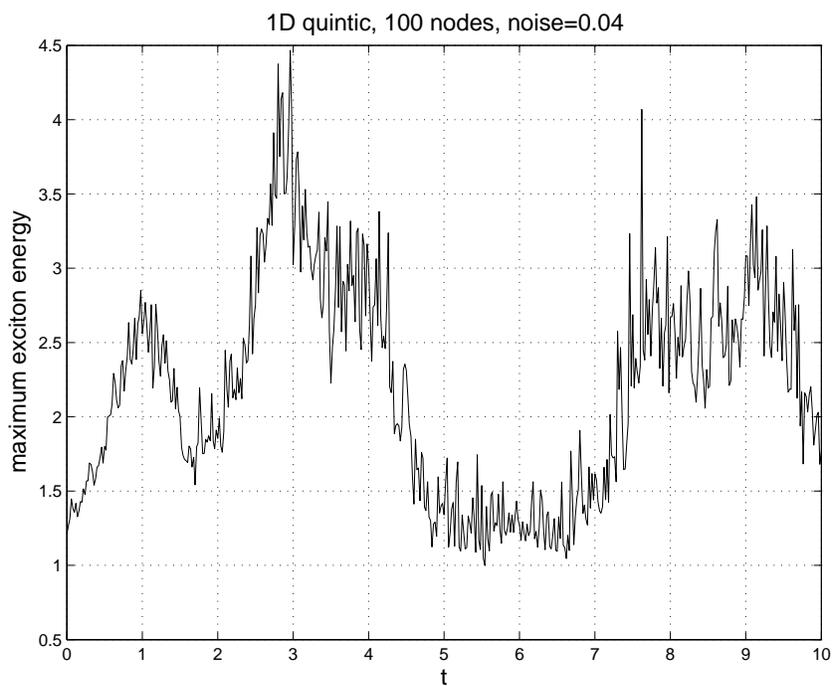}
\caption{Maximum single node energy for 1D quintic, noise $\gamma=0.04$, no damping.}
\label{fig:dnlsn_edmaxt}
\end{figure}

The evolution of the degree of localization, or lack thereof, is shown in Fig.~\ref{fig:dnlsn_edmaxt}.
Note that transient spikes to values more than four times the initial value occur, above all close to the focusing time $t^* \approx 2.6$ noted above, but these are not indications of focusing towards localization.
Continuing this solution for considerably more time never again reaches the maximum of about 4.5 seen at $t\approx 3$.

\subsubsection{Effects of adding both damping and noise}

\begin{figure}[htbp]
\includegraphics
[height=\graphicsheight]
{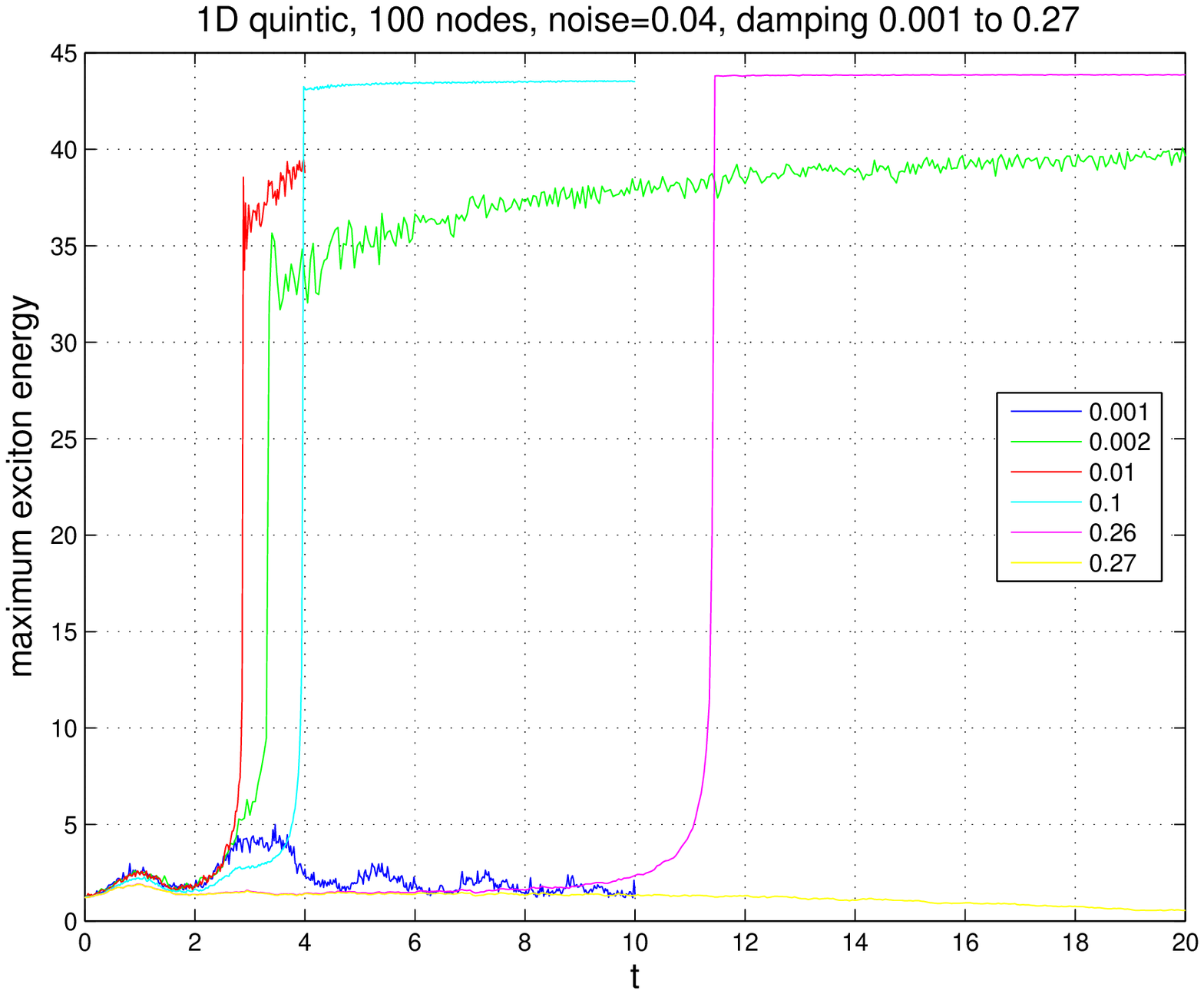}
\caption{Maximum single node energy for 1D quintic, noise $\gamma=0.04$,
damping from $\lambda=0.001$ to $0.27$.}
\label{fig:dnlsnd_edmaxt_dvarious}
\end{figure}

\begin{figure}[htbp]
\includegraphics
[height=\graphicsheight]
{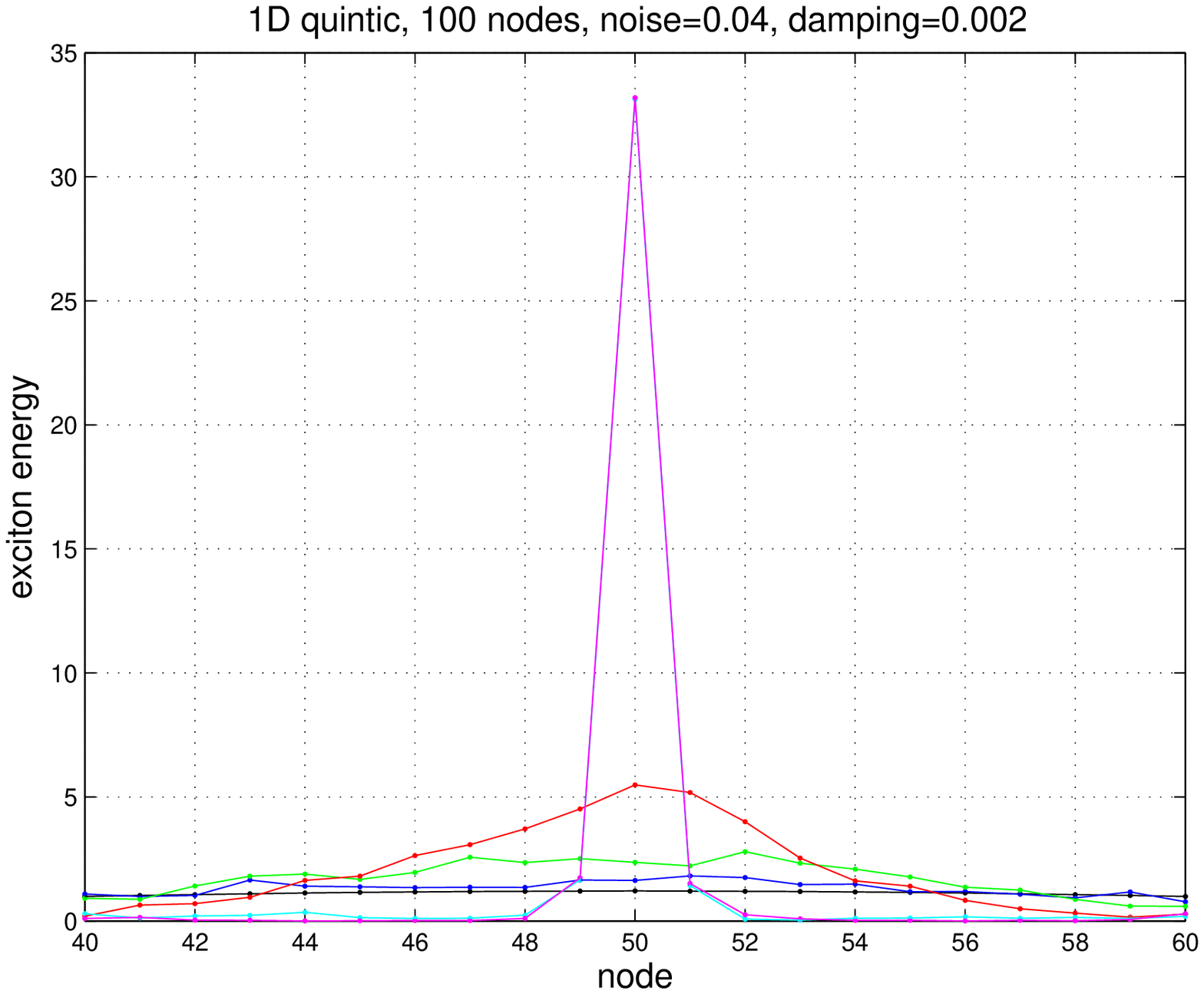}
\caption{Energy distribution for 1D quintic, noise $\gamma=0.04$, damping $\lambda=0.002$.}
\label{fig:dsdst1dq_n04_d2e_3_etx}
\end{figure}

The effect of adding damping at various strengths $\lambda=0.001, 0.002, 0.01, 0.1, 0.26, 0.27$ while maintaining the noise level of $\gamma=0.04$ is summarized in Fig.~\ref{fig:dnlsnd_edmaxt_dvarious} in terms of the maximum single node energy, and spatial structure is shown for the case $\lambda= 0.002$ in Fig.~\ref{fig:dsdst1dq_n04_d2e_3_etx}.

Small damping values ($\lambda \leq 0.001$) cause a transient increase in collapse, but not enough to produce localization; instead, one eventually gets dispersion, as with damped noise.
With damping above some threshold near $\lambda=0.002$, collapse proceeds to persistent localization of most energy at a single node.
With the larger damping values $\lambda=0.01,0.1$, the maximum single node energy quickly becomes almost constant at a value very close to 45.3, which as noted above means that almost all energy is at one node.

Apparently solutions settle into a very close approximation of a steady state that has energy almost completely localized.
This is also true at the lower damping values for which localization is seen, but with far slower onset.
For $\lambda=0.002$ maximum single node energy continues the rising trend seen, reaching 41 by $t=50$ and 43 by $t=100$.

\begin{figure}[htbp]
\includegraphics
[height=\graphicsheight]
{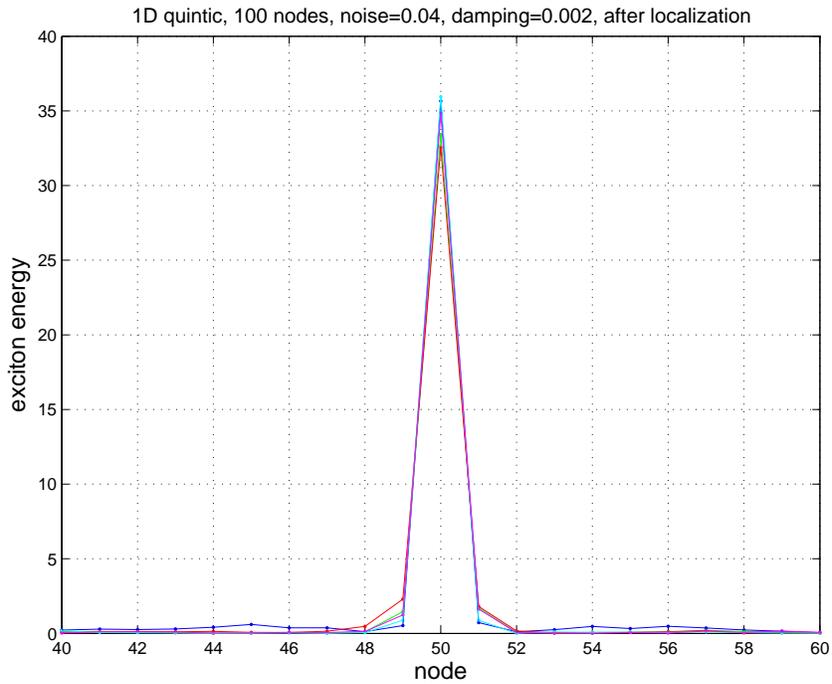}
\caption{As above but for times $t=3.4, 3.8, 4.2, 4.6, 5$ after the onset of energy localization, showing the persistence of a near steady state.}
\label{fig:dsdst1dq_n04_d2e_3_etx_after}
\end{figure}

This strong spatial localization even for $\lambda=0.002$ is shown in Fig.~\ref{fig:dsdst1dq_n04_d2e_3_etx_after}, which gives data for five times after onset of localization.
This and the previous graph show a form far closer to a steady state than for the undamped, noiseless case above.
The strong oscillations seen previously are absent here, replaced only by far smaller fluctuations, as are inevitably caused by the noise.

\begin{figure}[htbp]
\includegraphics
[height=\graphicsheight]
{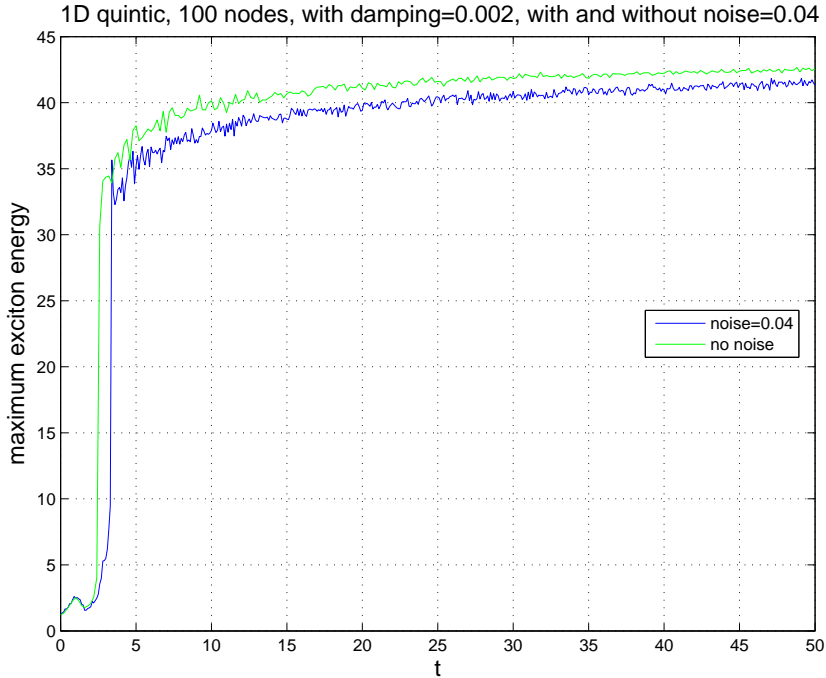}
\caption{Maximum single node energy for 1D quintic, damping $\lambda=0.002$, for both noise $\lambda=0.04$ and no noise.}
\label{fig:dnlsd_edmaxt}
\end{figure}

Fig.~\ref{fig:dnlsd_edmaxt} shows the approach to a nearly stationary state over a longer time interval as indicated by maximum single node energy.
It also gives the solution with the same damping but no noise: there is relatively little difference, indicating that damping is the dominant mechanism driving the solution towards a steady state, and that this mechanism is robust enough to be little perturbed by noise.

A final observation is that as damping strength $\lambda$ is increased, the localization initially occurs earlier, then later, and finally, localization completely fails with a sharp transition between $\lambda=0.26$ and 0.27.
Sufficiently strong damping apparently causes spatial smoothing which not only counteracts noise effects as before but also inhibits self-focusing.

\subsubsection{Connections to evolution of the Hamiltonian}
\label{subsubsect: numerical results, hamiltonian}

There are indications in Eq's (\ref{eq:variance-deriv-snls}-\ref{eq:dHdt-dsdnls}) that the evolution of the
Hamiltonian $\Ham$ could be related to the occurrence of self-focusing and localization of energy, as it is for the NLS, so this will be examined.

First, it can be shown that the result of Eq's~(\ref{eq:ham-deriv-snls},\ref{eq:R}) apply at least qualitatively to the discrete system with noise but no damping, by considering the latter as a discretization of the Stochastic NLS with noise having correlation length scale proportional to the lattice spacing $l$.
Since the precise constant of proportionality is not known, this will be done by checking first that $\langle \Ham \rangle$ grows roughly linearly in time, and then by observing that this linear growth rate is roughly proportional to $n_{nodes}^3$, as suggested by Eq.~\eqref{eq:R}.

\begin{figure}[htbp]
\includegraphics
[height=\graphicsheight]
{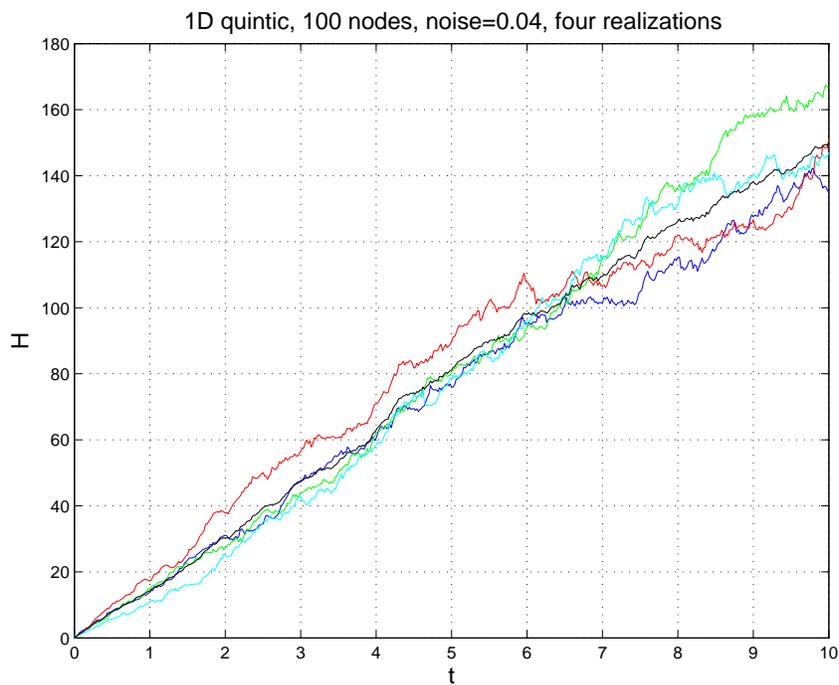}
\caption{Hamiltonian for 1D quintic, noise $\gamma=0.04$, no damping.
Colored curves are for different noise realizations, the black curve is their mean.}
\label{fig:1dq_dnlsn_Ht_m}
\end{figure}

The evolution of $\langle \Ham \rangle$ is approximated by the black curve in Fig.~\ref{fig:1dq_dnlsn_Ht_m}.
This is the average of results for four noise realizations shown by the colored curves, of which the blue curve corresponds to the single realization used in earlier graphs.
It is seen that there is indeed roughly linearly growth in time, at an average rate of about 15.

\begin{figure}[htbp]
\includegraphics
[height=\graphicsheight]
{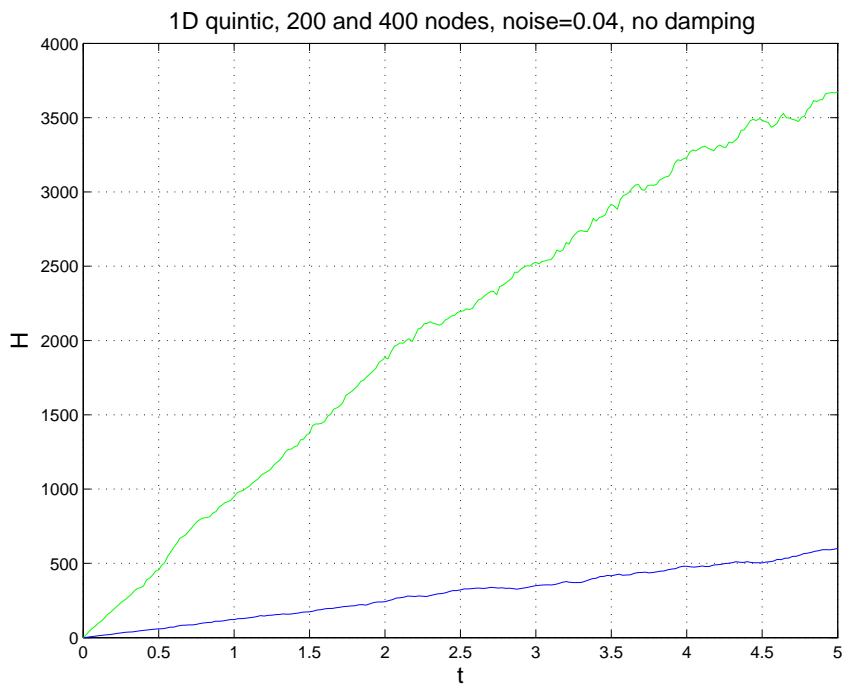}
\caption{Hamiltonian for 1D quintic, noise $\gamma=0.04$, no damping. Blue is with 200 nodes, red is with 400, one realization for each.}
\label{fig:1dq_dnlsnd_Ht_n200400}
\end{figure}

Increasing $n_{nodes}$ with initial data and noise adjusted as for grid refinement in discretization of the SNLS, the growth rate of the Hamiltonian is seen in Fig.~\ref{fig:1dq_dnlsnd_Ht_n200400}
to go from about 15 for $n_{nodes}=100$ to $100$--$120$  for $n_{nodes}=200$ and $740$--$1000$ for $n_{nodes}=400$.
Though only a single realization is shown in each case, the growth rates are again approximately linear, and the changes in the rates are consistent with the predicted $n_{nodes}^3$ scaling.
As noted in section \ref{subsect:background, localization in snls}, this indicates the ill-posedness of SNLS with spatially uncorrelated noise.

The addition of damping is predicted by Eq.~\eqref{eq:dHdt-dsdnls} to at least partially offset the growth of $\Ham$ caused by noise.
It is natural to ask under what circumstances this effect is sufficient to bring $\Ham$ back to negative values, and how the restoration of negative $\Ham$ is related to the restoration of energy localization, and the answers appear to be positive.

In the less interesting case of damping insufficient to restore energy localization, $\Ham$ initially grows roughly linearly as without damping, but then levels out to fluctuation around a significantly positive value.

\begin{figure}[htbp]
\includegraphics
[height=\graphicsheight]
{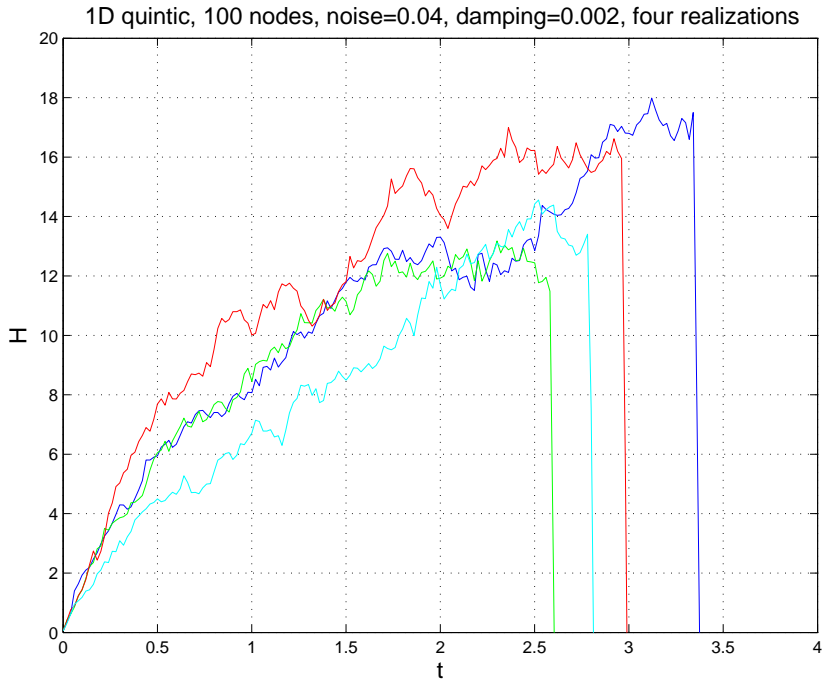}
\caption{Hamiltonian for 1D quintic,  noise $\gamma=0.04$, damping $\lambda=0.002$.
Same noise realizations and curve color-scheme as in Fig.~\ref{fig:1dq_dnlsn_Ht_m}.}
\label{fig:1dq_dnlsnd_Ht_m}
\end{figure}

\begin{figure}[htbp]
\includegraphics
[height=\graphicsheight]
{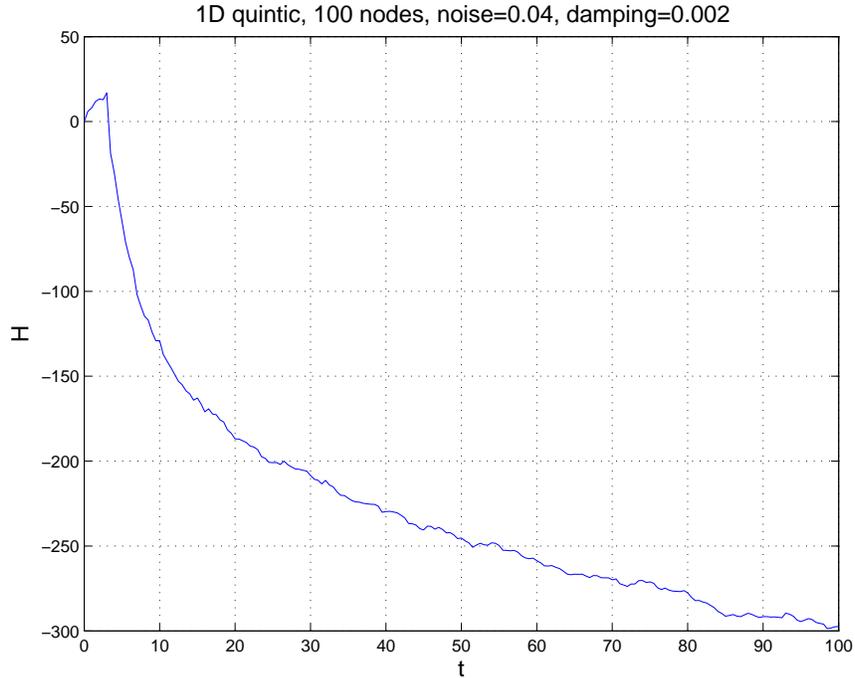}
\caption{The first realization above to $t=100$.}
\label{fig:1dqnn_ht_dnlsnd_long}
\end{figure}

With damping sufficient to restore energy localization ($\lambda=0.002$, $\gamma=0.04$),
Fig.~\ref{fig:1dq_dnlsnd_Ht_m} shows for each of four realizations there is initial roughly linear growth of $\Ham$ at about the same rate 15 seen above, but this is followed by slowing of the growth, and then a sudden drop to negative values.
These negative values quickly become very large and remain so, as shown for the standard noise realization in
Fig.~\ref{fig:1dqnn_ht_dnlsnd_long}.
For each realization, the time of this drop is fairly close to the time at which localization occurs, and indeed is driven by the sixth power nonlinearity term in $\Ham$ taking on a large negative value when the energy at any one node is a substantial proportion of the total.

The longer time behavior of  $\Ham$ shown in Fig.~\ref{fig:1dqnn_ht_dnlsnd_long} is continued decrease, correlated to the move closer to a localized steady state that is also indicated by Fig.~\ref{fig:dnlsd_edmaxt}.
The values of both maximum single node energy and $\Ham$ seem to settle slowly near asymptotic values, but with persistent noise driven fluctuations.
However, whereas maximum single node energy gets quite close to the extreme value $\N/l\approx 45.3$, the value of $\Ham$ does not get so close to the minimum possible value of
$\Ham_{min}=2(N-N^3/6)/l^2 \approx -511$, though both extrema occur for the same case of having all energy at one node.
The extra energy is presumably persistent spatial variations or thermal energy in the free energy part of $\Ham$.

\subsection{Discrete 2D radially symmetric DSNLS}

A nearest neighbor discretization of the 2D cubic DSNLS with radial symmetry has also been studied,
in order to test the conjectures described in section~\ref{subsect:background-localization-discrete}, based on a collective coordinates reduction of this 2D cubic case.
Another goal is computationally efficient initial comparison to simulations of the full 2D cubic model with noise but no damping by Bang et al \cite{BangCIRG1994}.

A weakness is the physically unlikely assumption of radially symmetric noise, but the qualitative features match those seen in the previous section, suggesting robustness under variations of lattice geometry, coupling, and nonlinearity power.

Indeed, no fundamental differences are seen from the previous study of $D=1$, $\sigma=2$ case, so we will simply summarize the points of agreement.

All results are for the initial data from discretization of $\psi(0,r) = 1.65 \mbox{sech}(r)$ on $0 \leq r \leq 10$ with 256 equally spaced nodes.
This function is close to the ground state ``Townes soliton'', giving initial value of the hamiltonean
$\Ham \approx -0.006$ and $\N \approx 1.8867$, just above the collapse threshold value
$N_c \approx 1.87$.
Thus singularity formation is ensured in the absence of noise and damping, but only just.
The idea is to maximize the sensitivity of self-focusing effects to the perturbations applied.

Noise without damping inhibits self-focusing localization of energy if its strength $\gamma$ exceeds a threshold, between $0.01$ and $0.02$.
The hamiltonian for each realization $\Ham$ grown roughly linearly in time, and more so the approximation of ensemble average $\langle\Ham\rangle$ given by averaging even a few realizations.
If the spatial discretization is refined to more nodes, the rate constant increases rapidly, suggesting the failure of the continuum limit and ill-posedness of this form of SNLS.

With above-threshold noise $\gamma=0.02$, damping restores energy localization once its strength $\lambda$ in turn exceeds a threshold, between $0.05$ and $0.1$.
This is related to first slower growth of $\Ham$, and then its decrease to substantially negative values at about the same time as the localization.

\subsection{1D lattice, quintic nonlinearity, spectral coupling}

Finally, a different spatial discretization has been used at the opposite extreme to nearest neighbor: discretization of the damped stochastic NLS equation for $D=1$, $\sigma=2$ with periodic boundary conditions as above, but with spectral discretization of derivatives.

The only changes seen are in quantitative details, not the qualitative features described for both previous cases, so details are omitted.

This coupling is unnatural for the underlying molecular systems due to a mixture of signs corresponding to alternation between attractive and repulsive dipole interactions.
However if continuum limits are valid for suitable combinations of noise and damping, so that there is smoothness on a scale larger than the molecular spacing, spectral discretization could potentially reduce computational costs.

\section{Refining the continuum limit approximations}
\label{sect:continuum}

As discussed above in sections \ref{subsect:background, localization in snls}, \ref{subsubsect:results, 1D Q, n, no d} and \ref{subsubsect: numerical results, hamiltonian}, the continuum limit from the undamped Stochastic Discrete NLS (Eq.~\ref{eq:DSDNLS} with $\lambda=0$) with spatially uncorrelated noise is probably ill-posed; a so-called ``ultra-violet catastrophe''.
Some brief comments on possible solutions are offered to indicate possible future research directions.

One possibility is that the addition of damping restores well-posedness, and numerical evidence above suggests that this might be so.
However, it is not clear that the damping term allows one to overcome the technical obstacles to establishing existence and uniqueness; instead its fully nonlinear term might only increase the technical difficulties.

A second approach is retaining higher order terms in approximating the discrete coupling terms by Taylor series expansions, which preserves explicit dependence on a length scale parameter $l$.

Assuming the Dihedral $D_2$ symmetry of the brick-wall molecular film structure, one gets
\begin{equation} \label{eq:J-expansion-D2}
 \sum_{m} J_{nm}\Psi_m = j_{2,0}\psi_{xx}+j_{0,2}\psi_{yy}
  + l^2[j_{4,0}\psi_{xxxx} + j_{2,2}\psi_{xxyy} + j_{0,4}\psi_{yyyy})] + O(l^4)
\end{equation}

With the physically natural assumption of non-trivial and attractive coupling, meaning that all  $J_{nm}$ are non-negative and some are positive, all the new $j_{ab}$ coefficients are non-negative and all $j_{a0},j_{0b}$ are positive.
Linear rescaling of the $x$ and $y$ variables can then give the form
\begin{equation} \label{eq:J-expansion-D4}
 \sum_{m} J_{nm}\Psi_m = \psi_{xx}+\psi_{yy} + O(l^2),
\end{equation}
and discarding  terms that vanish as $l \to 0$ lead to the standard DSNLS approximation in Eq.~\eqref{eq:DSNLS}.

If instead we retain the next most important terms, involving fourth order spatial derivatives, and assuming $D_2$ symmetry, one can rescale to get
\begin{equation} \label{eq:MDSNLS}
i \frac{\partial \psi}{\partial t}
+ \lap\psi
{ + l^2[j_{4,0}(\psi_{xxxx} + \psi_{yyyy}) + j_{2,2}\psi_{xxyy}]}
+ |\psi|^2\psi 
+ \gamma \sigma \psi
- \lambda \frac{\partial (|\psi|^2)}{\partial t} \psi
= 0.
\end{equation}
With nearest neighbor coupling on a rectangular lattice the cross derivative term vanishes ($j_{2,2}=0$), but that does not fit the brick-wall symmetry observed for cyanine dye Scheibe aggregates.

A third approach is to use Pad\'{e} fitting to the coupling term instead of Taylor polynomials, giving a pseudo-differential equation.
With nearest neighbor coupling on a rectangular lattice one can get
\begin{equation} \label{eq:MDSNLS-pade}
i \frac{\partial \psi}{\partial t}
{ + \left[I- \frac{l^2}{12}\lap\right]^{-1} \kern-1ex \lap\psi}
+ |\psi|^2\psi + \gamma \sigma \psi - \lambda \frac{\partial (|\psi|^2)}{\partial t} \psi
= 0.
\end{equation}
One potentially advantage is that the pseudo-differential operator is bounded, which explicit removes the risk of an ``ultra-violet catastrophe'' and so might facilitate analysis.

\section{Conclusions and Plans}

\begin{itemize}
\item
Noise without damping can inhibit wave collapse, related to an increase in the hamiltonian.
\item
There is a minimum threshold noise level needed to do this.
\item
The continuum limit as a Stochastic NLS equation with spatially uncorrelated noise appears to be ill-posed.
\item
Damping can restore collapse, again requiring a threshold strength.
\item
Damping might also restore well-posedness with spatially uncorrelated noise.
\item
Even more damping inhibits self-focusing and collapse.
\end{itemize}

In further work, the model should be refined in various ways.
\begin{itemize}
\item
Simulations and analysis of full 2D models.
\item
Consideration of the 1D cubic case and related models of nonlinear waves in long bio-molecules, where energy localization still occurs in discrete NLS models but the NLS continuum limit does not have self-focusing collapse.
\item
Analysis of well-posedness of the various PDE and pseudo-differential equation models, including the effects of damping.
\item
Direct analysis and study of the ODE system (lattice models) with various interaction forms such as longer range.
\item
Simulation with time correlation in the noise, as seen in Eq.~\eqref{eq:sigma-from-eta}.
\end{itemize}

\section*{Acknowledgements}

The first author gives thanks to Peter Christiansen and Yuri Gaididei for numerous discussions, to the IMM at the Danish Technical University for supporting several visits, and to the Department of Mathematics at the University of Arizona where he is currently visiting.

The authors was supported in part by the College of Charleston 4th Century Initiative through a Summer Undergraduate Research Grant.


\end{document}